\documentclass[11pt,twoside]{article}

\usepackage{asp2014}

\aspSuppressVolSlug
\resetcounters

\begin{document}

\title{Realtime alerts of the transient sky on mobile devices}

\author{P.~Reichherzer,$^{1,2,3}$ F.~Sch\"{u}ssler,$^1$ V.~Lefranc,$^1$ A.~Alkan,$^1$ and J.~Becker~Tjus$^{2,3}$}
\affil{\email{astro.colibri@gmail.com}}
\affil{$^1$\small IRFU, CEA, Universit\`{e} Paris-Saclay, F-91191 Gif-sur-Yvette, France}
\affil{$^2$\small Ruhr-Universit\"{a}t Bochum, Universit\"{a}tsstra{\ss}e 150, D-44801 Bochum, Germany}
\affil{$^3$\small Ruhr Astroparticle and Plasma Physics Center, Ruhr-Universit\"{a}t Bochum, D-44780 Bochum, Germany}

\paperauthor{P.~Reichherzer}{patrick.reichherzer@rub.de}{0000-0003-4513-8241}{Ruhr-Universit\"{a}t Bochum}{Theoretische Physik IV}{Bochum}{Nordrhein westfalen}{44801}{Germany}
\paperauthor{F.~Sch\"{u}ssler}{fabian.schussler@cea.fr}{0000-0003-1500-6571}{CEA}{IRFU}{Gif-sur-Yvette}{State/Province}{91191}{France}
\paperauthor{V.~Lefranc}{valentin.lefranc@cea.fr}{0000-0002-4806-8931}{CEA}{IRFU}{Gif-sur-Yvette}{State/Province}{91191}{France}
\paperauthor{A.~K.~Alkan}{atilla.alkan@cea.fr}{0000-0001-7964-4420}{CEA}{IRFU}{Gif-sur-Yvette}{State/Province}{91191}{France}
\paperauthor{J.~Becker~Tjus}{julia.tjus@rub.de}{0000-0002-1748-7367}{Ruhr-Universit\"{a}t Bochum}{Theoretische Physik IV}{Bochum}{Nordrhein westfalen}{44801}{Germany}



\begin{abstract}
Follow-up observations of transient events are crucial in multimessenger astronomy. We present Astro-COLIBRI as a tool that informs users about flaring events in real-time via push notifications on their mobile phones, thus contributing to enhanced coordination of follow-up observations.
We show the software's architecture that comprises a REST API, both a static and a real-time database, a cloud-based alert system, as well as a website\footnote{\url{https://www.astro-colibri.com}} and apps for iOS and Android as clients for users. The latter provide a graphical representation with a summary of the relevant data to allow for the fast identification of interesting phenomena along with an assessment of observing conditions at a large selection of observatories around the world in real-time.
\end{abstract}



\section{Transient alerts in multimessenger Astronomy}
The field of astrophysics is experiencing several fundamental changes, such as the increasing relevance of observations of transients, i.e., temporal variabilities of astronomical sources such as supernova explosions, fast radio bursts (FRBs), and gamma-ray bursts (GRBs). An increasing number of complementary cosmic messengers provide crucial information about these events. The detections of a diffuse flux of high-energy neutrinos and of gravitational waves from mergers of compact binaries have started to supplement electromagntic observations. Combining these complementary messengers can provide crucial insights on events in the transient sky. The study of these extreme (flaring) events in the multimessenger approach requires real-time and, to a large extent, fully automated communication between telescopes (see e.g.\,\citet{2019ICRC743M,2021JCAP...03..045A}) and researchers to respond instantaneously to events with follow-up observations to better understand the nature of these events through complementary observational data. To tackle these challenges, among others, an ecosystem of numerous complimentary services has been established over the last decades. For a review of high-energy alerts in the multimessenger era, see \citet{Dorner2021}.
\newpage

\section{Realtime alerts on mobile devices}
Whereas the time-critical decisions for follow-up observations must be automatic, especially the evaluation and the decision on subsequent observations require extensive processing and a graphical representation, easily comprehensible for humans, of the information available on the event in the context of already known sources and temporal and spatial correlations of other transient signals.
\articlefigure{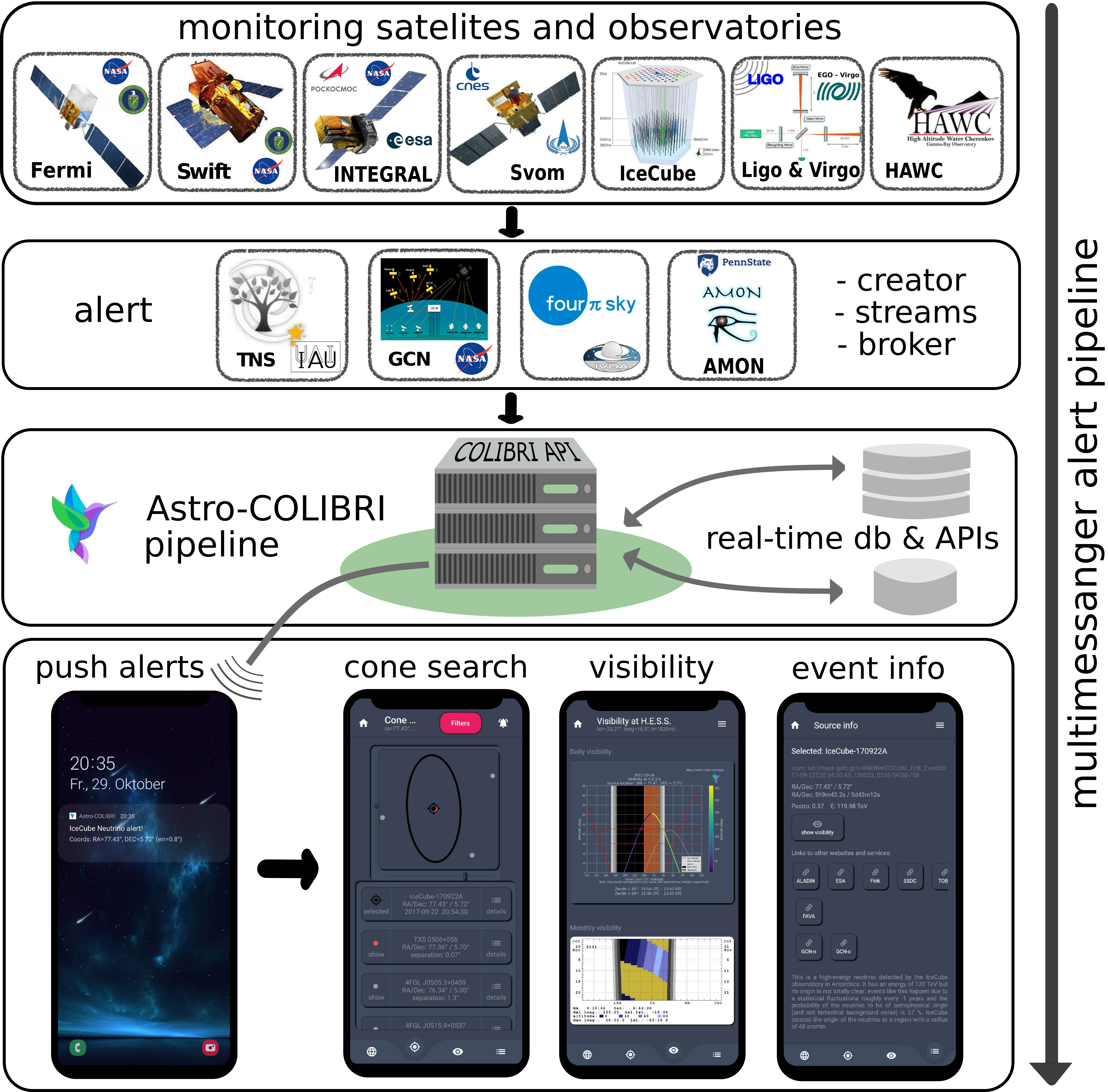}{fig1}{The multimessenger pipeline from observations of transient events to the user notification and its visualization in the Astro-COLIBRI app. Most processes are fully automatized and with negligible latency and downtime. Exceptions are e.g. the manual creation of GCN circulars that deliver information in human-written reports about the observed event. Details of the Astro-COLIBRI pipeline and the message transport of the push alerts to the mobile devices are shown in Fig.~\ref{fig2}.}

Astro-COLIBRI \citep{Reichherzer2021} aims to be the top layer that connects existing subsystems into a large ecosystem, with a focus on optimized display for users through an interactive graphical user interface (GUI) and mobile apps that receive push notifications in real time, all usable in an intuitive way. The multimessenger pipeline from observations to the user is discussed in the following and sceteched in Fig.~\ref{fig1}.
\begin{enumerate}
    \item \textit{Monitoring satellites and observatories:} The foundation of multimessenger astronomy are satellites and observatories that constantly monitor the various messengers of the transient sky. Gravitational waves (Ligo and Virgo), photons (Fermi, HAWC, Swift, INTEGRAL, SVOM, etc.), as well as neutrinos (IceCube, ANTARES, and GVD), can be localized very rapidly, which makes them essential for follow-up observations. Alerts announcing new detections are emitted largely automatically when specific criteria are met.
    \item \textit{Alert creators / streams / broker:} Alert and correlation systems such as VOEvent alerts, Astrophysical Multimessenger Observatory Network alerts \citep{AMON_2020}, Transient Name Server (TNS) notifications\footnote{\url{https://www.wis-tns.org/}}, Gamma-ray Coordinates Network (GCN) circulars\footnote{\url{https://gcn.gsfc.nasa.gov/}}, brokers (e.g. via 4 Pi Sky; \citet{4pisky_2016}, FINK broker; \citet{FINK_2021}) provide alerts which differ both in their degree of machine or human-generated content and in the underlying communication technology. The technologies involve the use of web pages, emails, and brokers and require the end-user to make an effort to keep track of the (ever-increasing amount of) relevant information.
    \item \textit{Astro-COLIBRI pipeline:} The Astro-COLIBRI pipeline processes machine or human-generated alerts in real-time and provides the event in the context of persistent sources as well as transient events in the relevant phase space. To accommodate the growing range of services from the community, Astro-COLIBRI is modular in design and allows easy integration of new or modifying third-party services as shown in Fig.~\ref{fig2}. 
    \item \textit{Push notifications:} Astro-COLIBRI informs users based on their preferences set in their Astro-COLIBRI app about these events in real-time (latency of a few 100\,ms) via push notifications on smartphones. Once opened the app, the interface provides a graphical representation with a summary of the relevant data to allow for the fast identification of interesting phenomena along with an assessment of observing conditions at a large selection of observatories around the world. See \citet{Schuessler2021} for a description of some use cases.
\end{enumerate}
\articlefigure{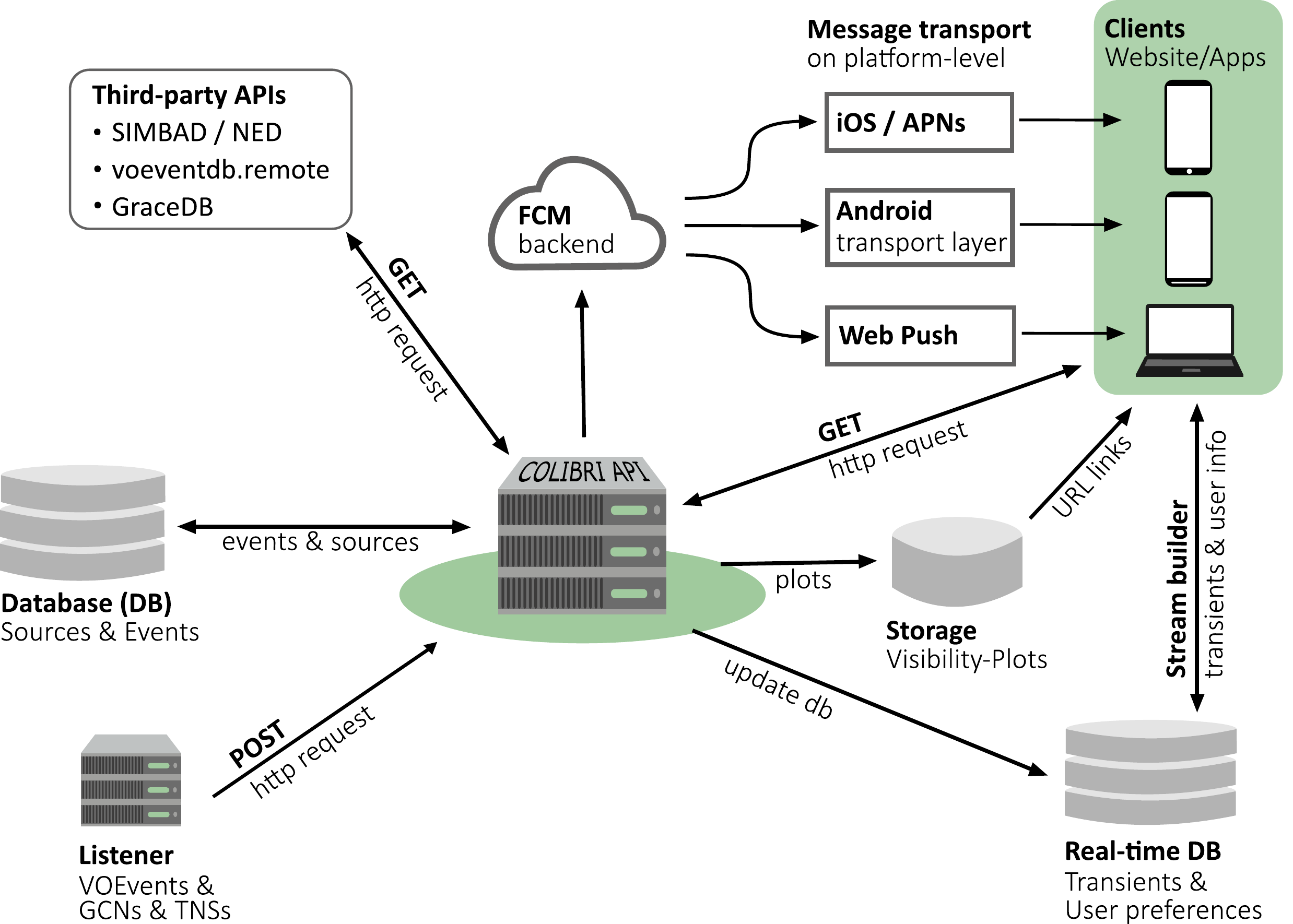}{fig2}{Architecture of Astro-COLIBRI comprises a REST API, both a static and a real-time database, a cloud-based alert system, as well as a website and apps for iOS and Android as clients. Public user interfaces are highlighted in green.}

\section{Outlook}
To enable a better exchange of relevant information on specific transient events, a comment function on individual events is planned. The collected information from the contributed comments can later be automatically processed and used in the context of new alerts. The automated evaluation of human-generated alerts will be accomplished in the future by means of deep learning-based natural language processing techniques within Astro-COLIBRI, thus enabling rapid extraction and synthesis of information from reports that will be associated with the respective transient events.\\\\
The Astro-COLIBRI development team welcomes comments and feedback from the community to further improve the platform. 
\newpage 
\acknowledgements This work was supported by the European Union's Horizon 2020 Programme under the AHEAD2020 project (grant agreement No. 871158). This work is supported by the ADI 2019 project funded by the IDEX Paris-Saclay, ANR-11-IDEX-0003-02 (PR). PR also acknowledges support from the German Academic Exchange Service and by the RUB Research School via the \textit{Project International} funding. JBT and PR acknowledge funding by the BMBF, grant number 05A20PC1.

\textit{Used software / packages / frameworks:} 4 Pi Sky~\citep{4pisky_2016}, Astropy \citep{2018AJ....156..123A}, Flutter\footnote{\url{https://flutter.dev/}}

\bibliography{X3-001}  

\end{document}